\documentclass{moriond}

\def\be{\begin{equation}}
\def\ee{\end{equation}}
\def\bea{\begin{eqnarray}}
\def\eea{\end{eqnarray}}

\newcommand{\Photo}{\includegraphics[height=35mm]{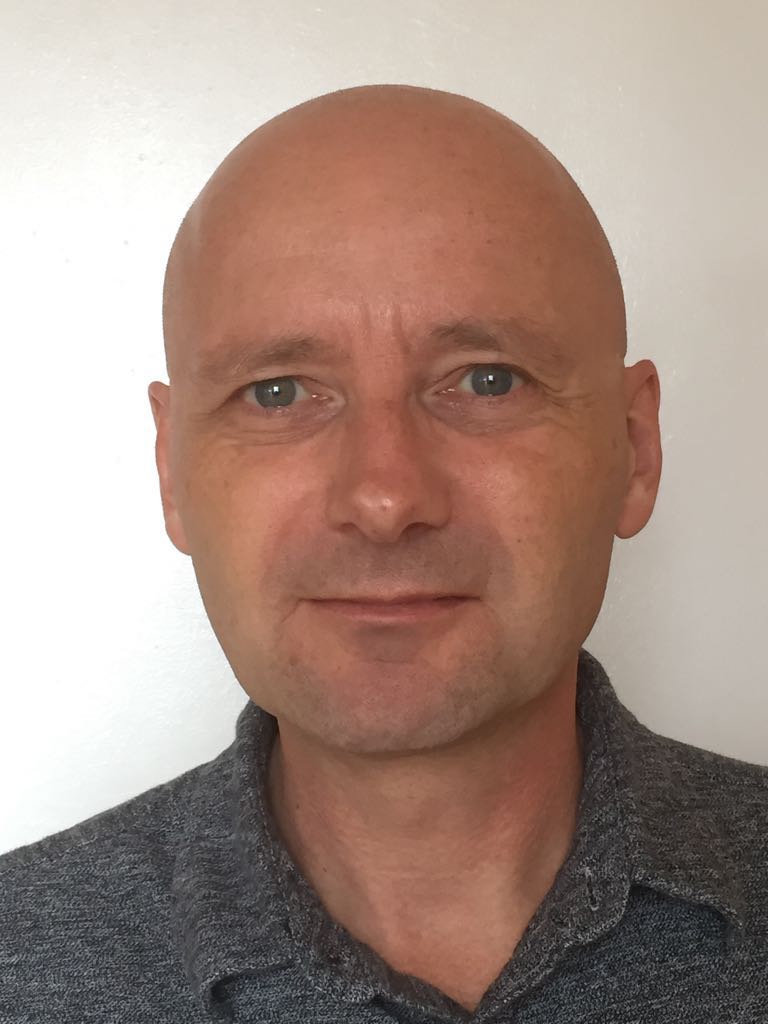}}
\usepackage[numbers]{natbib}
\usepackage{amsmath,amssymb}

\begin{document}
\vspace*{4cm}
\title{PRECISION ELECTROWEAK MEASUREMENTS AT RUN 2 AND BEYOND}

\author{JENS ERLER}

\address{Instituto de F\'isica, Universidad Nacional Aut\'onoma de M\'exico, \\
Apartado Postal 20--364, CDMX 01000, M\'exico\vspace*{12pt}}

\maketitle\abstracts{
After reviewing the key features of the global electroweak fit, I will provide updated results
and offer experimental and theoretical contexts. 
I will also make the case for greater precision and highlight future directions.}

\section{Introduction}
To chase out the elephants in the room, I recall that with the Higgs boson discovery the Standard Model (SM) is now complete,
and with very few marginal exceptions it passed all the tests.
Furthermore, the LHC did not yet find any convincing evidence for physics beyond the SM.
Nevertheless, if nothing else does, at least dark matter provides a solid observational hint at the presence of new physics,
and it may quite plausibly linger near the electroweak (EW) scale. 
Perhaps we are witnessing a revival of the times where precision physics is guiding high energy physics, like in the era of LEP.
It could be that the renormalizable SM is merely the leading set of terms in a non-renormalizable effective quantum field theory,
where the former gives rise to the (relatively) long-range physics. 

Here I review and update the global electroweak (EW) fit, restricting myself to the CP-even and flavor-diagonal part of the SM.
For more flavorful observables I refer to the contribution by Jure Zupan~\cite{Zupan}.
I will also allow certain model-independent parameters describing new physics.

\section{Precise inputs for the electroweak fit}
\subsection{Bosonic sector}
\label{subsec:EWfit}
The EW fit needs five input variables to define the bosonic sector of the SM,
namely the three gauge couplings associated with $SU(3)_C \times SU(2)_L \times U1)_Y$
and the two parameters entering the Higgs potential.
It is inessential which parameters or observables we call inputs and which ones output
because there is no fundamental distinction between those in a global fit.
Nevertheless, one may think of the most precise ones as inputs parameters and these are listed in Table~\ref{tab:bosonicinputs}.

\begin{table}[h]
\caption[]{Convenient set of input parameters to fix the bosonic sector of the SM.
While the most precise value of $\alpha$ in the Thomson limit currently derives from the anomalous magnetic moment of the electron, 
$g_e - 2$, we list here the value extracted from the Rydberg constant, saving $g_e - 2$ as an additional {\em derived\/} observable 
which can then be employed to constrain certain types of new physics models.
$G_F$ is calculated using the measured muon lifetime. 
$M_Z$ is an output of the $Z$ line-shape fit at LEP~1.
$M_H$ is the result of the kinematic event reconstruction at the LHC and comparatively less precise,
but except for the total Higgs width it enters only in loops.
The value of $\alpha_s(M_Z)$ is from the global electroweak fit and dominated by $Z$ and $\tau$ decay observables.}
\label{tab:bosonicinputs}
\vspace{0.4cm}
\begin{center}
\begin{tabular}{|c|c|c|c|c|}
\hline
quantity & quoted as & central value & relative error & reference \\ \hline
fine structure constant & $\alpha^{-1}$ & 137.035999037 & $6.6 \times 10^{-10}$ & \cite{Bouchendira:2010es} \\
Fermi constant  & $(\sqrt{2} G_F)^{-1/2}$ & 246.21965~GeV & $2.6 \times 10^{-7}$ & \cite{Webber:2010zf,PDG2016} \\
$Z$ boson mass & $M_Z$ & 91.1876~GeV & $2.3 \times 10^{-5}$ & \cite{ALEPH:2005ab} \\
Higgs boson mass & $M_H$ & 125.09~GeV & $1.9 \times 10^{-3}$ & \cite{Oda} \\
strong coupling constant & $\alpha_s(M_Z)$ & 0.1182 & $1.4 \times 10^{-2}$ & \cite{PDG2016} \\ \hline
\end{tabular}
\end{center}
\end{table}

\subsection{Top quark mass}
Greater precision in the top quark mass, $m_t$, still matters in EW fits.
Indeed, the small change from the value used about 18~months ago~\cite{PDG2016}, $m_t = 173.34 \pm 0.81$~GeV,
reduces the fitted Higgs boson mass by about 3~GeV.
Very recently, ATLAS~\cite{Owen}, CMS~\cite{Owen}, and the Tevatron EW Working Group~\cite{Bartos},
each released combinations of their various top quark mass determinations. 
The results are listed in Table~\ref{tab:mt}.
For the grand average, needed in the fits below, 
I assumed that there is a systematic uncertainty of 0.29~GeV that is common among all three.
It is the sum (in quadrature) of the error components induced by the Monte Carlo generator, parton distribution functions, and QCD,
as obtained by ATLAS.
For comparison, the Tevatron modeling plus theory error amounts to 0.38~GeV and the CMS modeling error is 0.41~GeV.
Other uncertainties are assumed uncorrelated between collaborations.
Notice, that the statistical precision\,\footnote{Precision is defined as the inverse of the square of the uncertainty.} 
of the grand average is not simply the sum of the statistical precisions of the individual combinations, as is sometimes assumed. 
Rather, the procedure developed in Ref.~\cite{Erler:2015nsa} should be applied.

\begin{table}[h]
\caption[]{Recent combinations of top quark mass measurements by ATLAS~\cite{Owen}, CMS~\cite{Owen}, 
and at the Tevatron~\cite{Bartos}.
The grand average (see the main text) of the three combinations is also shown.
Despite appearances (due to delicate round-offs) the total (experimental) error of the grand average 
is the sum in quadrature of its statistical and systematic components.  
All entries are in GeV.}
\label{tab:mt}
\vspace{0.4cm}
\begin{center}
\begin{tabular}{|c|c|c|c|c|}
\hline
& central value & statistical error & systematic error & total error \\ \hline
ATLAS &172.84 & 0.34 & 0.61 & 0.70 \\
Tevatron & 174.30 & 0.35 & 0.54 & 0.64 \\ 
CMS & 172.43 & 0.13 & 0.46 & 0.48 \\ \hline
grand average & 172.97 & 0.13 & 0.38 & 0.41 \\ \hline
\end{tabular}
\end{center}
\end{table}

To the total experimental error one has to add a common theory error because the quoted values are {\em interpreted\/} 
to either represent  the top quark pole mass, $m_t$, or some other operational mass definition supposedly coinciding 
with the pole mass roughly within the strong interaction scale $\Lambda_{\rm QCD}$ (taken here as 500~MeV).
Thus, the constraint used in the fits is 
\be
m_t = 172.97 \pm 0.28_{\rm uncorr.} \pm 0.29_{\rm corr.} \pm 0.50_{\rm theory} \mbox{ GeV} = 172.97 \pm 0.64 \mbox{ GeV},
\label{eq:mtworld}
\ee
where I have split the experimental error into uncorrelated and correlated components.
The uncertainty of ${\cal O}(\Lambda_{\rm QCD})$ is assumed to also account for the uncertainty in the relation between
the top quark pole and $\overline{\rm MS}$ mass definitions.
By accounting for the leading renormalon contribution in this relation, 
it may ultimately be possible to reduce this uncertainty to about 70~MeV~\cite{Beneke:2016cbu}.

\subsection{Charm and bottom quark masses}
I should mention the increasing importance the charm and bottom quark masses, $m_c$ and $m_b$, have on the EW fit.
If they are known very precisely, one can use perturbative QCD to calculate the heavy quark contributions
to the renormalization group evolution of $\alpha$ from the Thomson limit to the $Z$ pole~\cite{Erler:1998sy},
and conversely of the weak mixing angle which has been measured precisely near the $Z$ pole (see Sec.~\ref{sec:onpole})
to lower energies~\cite{Erler:2004in}.

Similarly, $m_c$ and $m_b$ enter the SM prediction~\cite{Erler:2000nx} of the anomalous magnetic moment of the muon, $g_\mu - 2$.
While I do not cover it here, I recall that $g_\mu - 2$ deviates by more than 4~standard deviations
if one includes $\tau$ decay spectral functions corrected for $\gamma$--$\rho$ mixing~\cite{Jegerlehner:2011ti}.
The latter brings $\tau$ decays into agreement with $e^+ e^-$ annihilation and radiative return data. 
Even though the charm quark is technically decoupling, its numerical effect enters at the same level into $g_\mu - 2$
as the hadronic light-by-light contribution, and an uncertainty of 70~MeV in $m_c$ would induce an error
comparable to the anticipated uncertainties in upcoming experiments at Fermilab and J--PARC.
Thus, one would like to know $m_c$ an order of magnitude more accurately than this.

Finally, the linear relationship~\cite{Haisch} between Higgs couplings and masses of the particles in the single Higgs doublet SM
can be studied precisely at future lepton colliders.
To match the projections of the charm and bottom Yukawa coupling measurements from the corresponding Higgs branching ratios
one needs knowledge of $m_c$ and $m_b$ to 8~MeV and 9~MeV, respectively.
Interestingly, Ref.~\cite{Erler:2016atg} calibrated the $m_c$ uncertainty in the first-principle relativistic QCD sum rule approach 
and fortuitously found the minimally required 8~MeV accuracy
(not accounting for the parametric uncertainty induced by $\alpha_s$ which should become negligible in the future).

\begin{figure}[t]
\centerline{\includegraphics[width=0.7\linewidth]{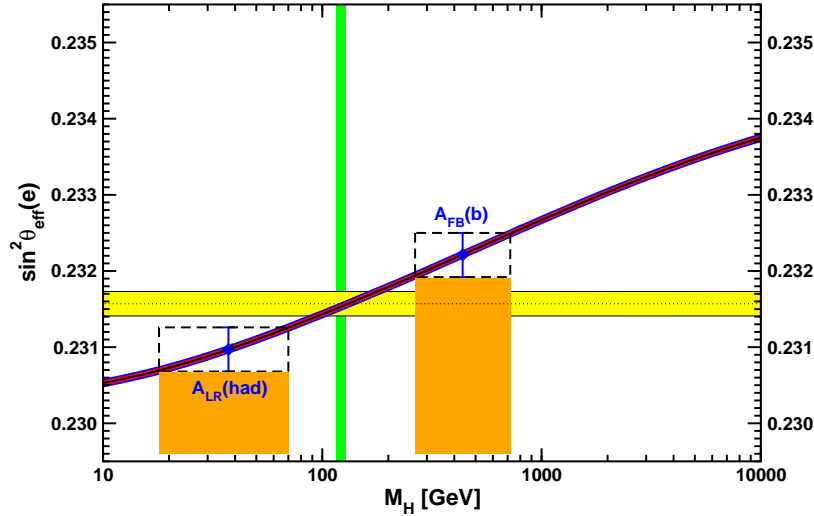}}
\caption[]{The left-right polarization asymmetry at the SLC favors values of $M_H$ which are too low and excluded,
while the forward-backward asymmetry for $b\bar b$ quark final states from LEP prefers Higgs boson masses of 300~GeV or more, 
which are also excluded.
It is only the average which actually agrees with the directly observed $M_H$.}
\label{fig:LEPSLCconflict}
\end{figure}

\section{The weak mixing angle}
\subsection{High-energy measurements}
\label{sec:onpole}
The weak mixing angle, $\sin^2\theta_W$, is one of two observables at the heart of the EW fit.
As a derived quantity, the strategy is to compute it and to compare it with $Z$ pole asymmetry measurements 
at LEP, the Tevatron and the LHC, from which the effective weak mixing angle for leptons, $\sin^2\theta_W^{\rm eff}$, is obtained.
An important application is to models with extra $Z^\prime$ bosons, 
in which $\sin^2\theta_W$ constrains the $Z$--$Z^\prime$ mixing angle typically to the $10^{-2}$ level~\cite{Erler:2009jh}.
The hadron collider measurements shown in Table~\ref{tab:sin2eff} agree well with each other,
but the two most precise $Z$ pole determinations are deviating by about 3~standard deviations
as illustrated in Fig.~\ref{fig:LEPSLCconflict}.

\begin{table}[h]
\caption[]{Measurements of $\sin^2\theta_W^{\rm eff}$ at the LHC~\cite{Andari} and the recent Tevatron combination~\cite{Han}.
The LHC average to be used in the fits is computed assuming that the smallest theoretical uncertainty 
($\pm 0.00056$ from LHCb) is fully correlated among the three LHC experiments~\cite{PDG2016}.}
\label{tab:sin2eff}
\vspace{0.4cm}
\begin{center}
\begin{tabular}{|c|c|c|c|c|}
\hline
& central value & statistical error & systematic error & total error \\ \hline
ATLAS ($\mu$ and $e$) & 0.2308 & 0.0005 & 0.0011 & 0.0012 \\
CMS ($\mu$) & 0.2287 & 0.0020 & 0.0025 & 0.0032  \\
LHCb ($\mu$) & 0.23142 & 0.00073 & 0.00076 & 0.00106 \\ \hline
LHC & 0.23105 & 0.00046 & 0.00074 & 0.00087 \\
Tevatron & 0.23179 & 0.00030 & 0.00017 & 0.00035 \\ \hline
\end{tabular}
\end{center}
\end{table}

\begin{figure}[t]
\centerline{\includegraphics[width=0.7\linewidth]{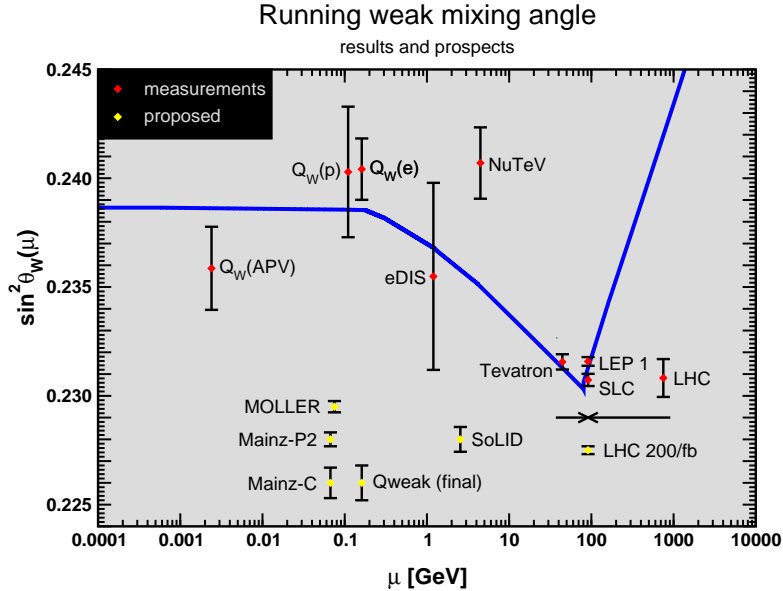}}
\caption[]{Current and future measurements of the weak mixing angle in the MS-bar scheme 
as a function of scale~$\mu$~\cite{Erler:2004in}.
The upcoming MOLLER experiment~\cite{Benesch:2014bas} in polarized M\o ller scattering at Jefferson Laboratory (JLab)
may almost reach the precision of the $Z$ factories LEP~1 and the SLC.
JLab also hosts the analogous Qweak experiment~\cite{Androic:2013rhu}
determining the left-right polarization asymmetry in elastic electron-proton scattering.
The P2 project~\cite{Bucoveanu:2016bgx} at the MESA facility under construction at the University of Mainz in Germany
is an even lower energy variant of Qweak.
There are further efforts, such as in parity-violating deep-inelastic scattering (PVDIS) 
at JLab using the SoLID detector~\cite{Souder:2016xcn}.
Also shown is a recent projection for Run~2 at the LHC~\cite{Bodek:2015bya}.}
\label{fig:s2w}
\end{figure}

\subsection{Low-energy measurements}
\label{sec:LE}
One can also compare the measurements of $\sin^2\theta_W$ near the $Z$ pole with off-pole determinations (see Fig.~\ref{fig:s2w})
to isolate possible new contact interactions.
This works because any four-fermion operator would be almost hopelessly suppressed under the $Z$ resonance,
but off the pole --- one could go to higher energies as well, but there are much more precise data at lower energies ---
there is a milder power suppression. 
Thus, if a significant difference between on-pole and off-pole measurements of $\sin^2\theta_W^{\rm eff}$ is observed 
it may be due to an effective contact interaction induced by TeV scale new physics.

\begin{figure}[t]
\centerline{\includegraphics[width=0.7\linewidth]{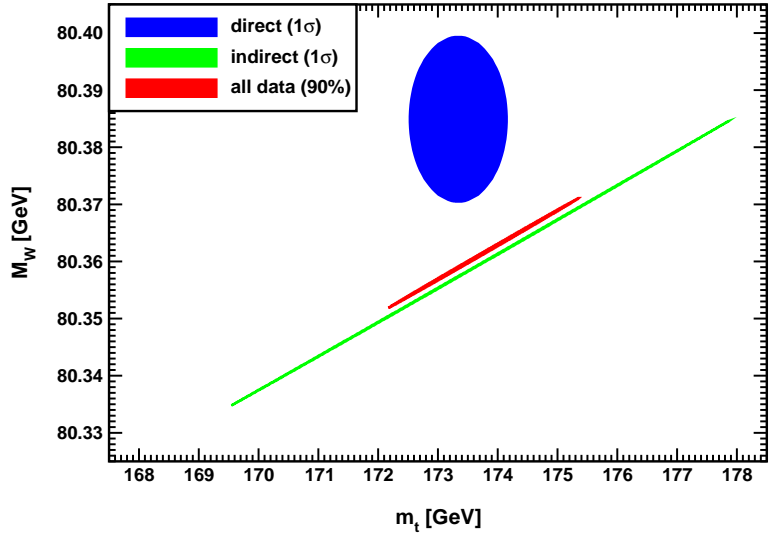}}
\caption[]{Direct and indirect determinations of the $W$ boson and top quark masses and the fit to all data~\cite{PDG2016}.}
\label{fig:mwmt}
\end{figure}

\section{Boson masses}
\subsection{W boson mass}
The other observable at the heart of the EW fit is the $W$ boson mass, $M_W$.
Its measurements at LEP~2 average to $M_W = 80.376 \pm 0.033$~GeV~\cite{Schael:2013ita},
while the Tevatron combination yields $M_W = 80.387 \pm 0.016$~GeV~\cite{Andari}.
The ATLAS result, $M_W = 80.3695 \pm 0.0185$~GeV~\cite{Andari}, represents the first at the LHC, 
and while it is based on only $4.6$~fb$^{-1}$ of 7~TeV data, 
it is already at the level of the most precise result at the Tevatron.
For what follows I assume a common PDF error of 7~MeV between the Tevatron and ATLAS uncertainties 
and will work with the average,  
\be
M_W = 80.379 \pm 0.012~{\rm GeV}  \qquad\qquad \mbox{(world average)},
\ee
corresponding to the weak mixing angle in the on-shell scheme,
\be
\sin^2\theta_W^{\rm on-shell} \equiv 1 - \frac{M_W^2}{M_Z^2} = 0.22301 \pm 0.00023\ .
\ee
However, the physics of $M_W$ and $\sin^2\theta_W$ is quite different, and while it is popular 
to convert one into the other, this is a fairly pointless exercise, especially in the context of new physics.
Rather, the two observables are complementary and provide, for example, 
constraints on the oblique parameters (see Sec~\ref{sec:oblique}) that are linearly independent.
The global fit returns
\be
M_W = 80.362 \pm 0.005~{\rm GeV} \qquad\qquad \mbox{(global fit)},
\ee
which is now driven by the directly measured $M_H$ and somewhat lower than the world average.

$M_W$ is of interest as it is easily affected by new physics in general and Higgs sector modifications in particular, 
but it needs the top quark mass, $m_t$, as an input. 
Fig.~\ref{fig:mwmt} compares the direct measurements of $M_W$ and $m_t$ from the colliders
to everything else in precision EW physics, including the direct value of $M_H$.
It is quite interesting that the measured $M_W$ is somewhat high,
because most kinds of new physics models addressing the EW hierarchy problem can easily affect the $M_W$ prediction. 
This includes the Minimal Supersymmetric Standard Model, where the shift in $M_W$ is predicted to be positive 
throughout parameter space~\cite{Heinemeyer:2013dia}, in agreement with what is currently favored by the data.

%\begin{figure}[t]
%\centerline{\includegraphics[width=0.7\linewidth]{mhmt_PDG_2016.pdf}}
%\caption[]{}
%\label{fig:mhmt}
%\end{figure}

\subsection{Higgs boson mass}
\label{sec:mh}
There are three different methods to determine $M_H$.
One employs Higgs boson branching ratios~\cite{PDG2016},
since especially the branching fractions into pairs of gauge boson feature a strong $M_H$ dependence. 
Using furthermore ratios of branching ratios, such as ${\cal B}(H \to \gamma\gamma)$ relative to 
${\cal B}(H \to WW)$ or ${\cal B}(H \to ZZ)$, cancels the dominant production uncertainties,
and we find~\cite{PDG2016},
\be
M_H = 126.1 \pm 1.9  \mbox{ GeV} \qquad\qquad \mbox{(branching ratios)}.
\ee
The global EW fit including updates presented at this meeting favors the rather low range,
\be
M_H = 94^{+18}_{-16} \mbox{ GeV} \qquad\qquad \mbox{(global fit)}.
\label{eq:mhfit}
\ee
This is about 1.7~$\sigma$ below the direct kinematic reconstruction result~\cite{Oda},
\be
M_H = 125.09 \pm 0.24 \mbox{ GeV} \qquad\qquad \mbox{(direct)}.
\ee
Thus, while $M_H$ is now known, it still provides a very valuable cross-check of the SM.

Before discussing the prospects at future LHC runs, 
it is entertaining to review how previous experimental projections compare with the actual achievements. 
Table~\ref{tab:snowmass} shows projections~\cite{Baur:2002gp} at the time of the Snowmass~2001 gathering 
on what was then thought to be the future of high-energy physics.
As one can see at the example of the Tevatron, with less than the expected integrated luminosity the goals were either achieved or 
surpassed and the finalized uncertainties may well turn out to be smaller, yet.
Similarly, the uncertainty of the first measurement of $M_W$ at the LHC with only one detector
and just a few~fb$^{-1}$ of data is approaching the 100~fb$^{-1}$ projection.
And $m_t$ from the LHC is already more accurate than anticipated.

\begin{table}[h]
\caption[]{Projections made in 2001 for various high-energy colliders.
For the LHC and Run~IIB at the Tevatron the numbers in parentheses show the currently achieved uncertainties.
The $\delta\sin^2\theta_W^{\rm eff}$ entry for the linear collider (LC) assumed a polarized fixed-target experiment 
analogous to the planned MOLLER experiment~\cite{Benesch:2014bas} at JLab, using the electron arm of the LC.
GigaZ refers to the $Z$ factory option at the LC.}
\label{tab:snowmass}
\vspace{0.4cm}
\begin{center}
\begin{tabular}{|c||c||c|c|c||c|c|}
\hline
& 2001 &Tev.~Run~IIA & Tev.~Run~IIB & LHC & LC & GigaZ \\
$\int {\cal L}$ [fb$^{-1}$] & --- & 2 & 15 (10) & 100 (30) & 500 & ---  \\ \hline
$\delta\sin^2\theta_W^{\rm eff}~(\times 10^5)$ & 17 & 78 & 29 (35) & 14--20 (87) & (6) & 1.3 \\
$\delta M_W$ [MeV] & 33 & 27 & 16 (16) & 15 (19) & 10 & 7 \\ 
$\delta m_t$ [GeV] & 5.1 & 2.7 & 1.4 (0.64) & 1.0 (0.5) & 0.2 & 0.13 \\ 
$\delta M_H$ [MeV] & --- & --- & ${\cal O}(2000)$ (---) & 100 (240) & 50 & 50 \\ \hline
\end{tabular}
\end{center}
\end{table}

The result in Eq.~(\ref{eq:mhfit}) is dominated by $M_W$, which by itself corresponds to $M_H = 89^{+22}_{-19}$~GeV.
A hypothetical measurement of $M_W = 80.376 \pm 0.008$~GeV (the assumed central value is adjusted 
so as to reproduce the current best fit value for $M_H$ and the error is motivated by Ref.~\cite{Bozzi:2015zja}) 
at the LHC after the accumulation of 150~fb$^{-1}$ of data yields $M_H = 94^{+17}_{-15} \mbox{ GeV}$. 
For this I assumed that the total $m_t$ error will be completely dominated by the QCD uncertainty in Eq.~(\ref{eq:mtworld}). 
And I neglected the theoretical error in the prediction of $M_W$, but to compensate I did not assume any improvement in other parameters 
like $\alpha_s$ or the electromagnetic coupling at the $Z$ scale.
Similarly, the hypothetical result $\sin^2\theta_W^{\rm eff} = 0.23135 \pm 0.00020$~\cite{Bodek:2015bya} 
would yield $M_H = 94^{+47}_{-32} \mbox{ GeV}$.
Adding these improvements to the current data gives $M_H = 90^{+13}_{-12} \mbox{ GeV}$.
Finally, at the high-luminosity LHC (HL--LHC) the uncertainty in $M_W$ may optimistically be reduced to 5~MeV,
and the one in $\sin^2\theta_W^{\rm eff}$ to $1.4 \times 10^{-4}$, which would then result in $M_H = 89 \pm 10$~GeV.

\begin{figure}[t]
\centerline{\includegraphics[width=0.7\linewidth]{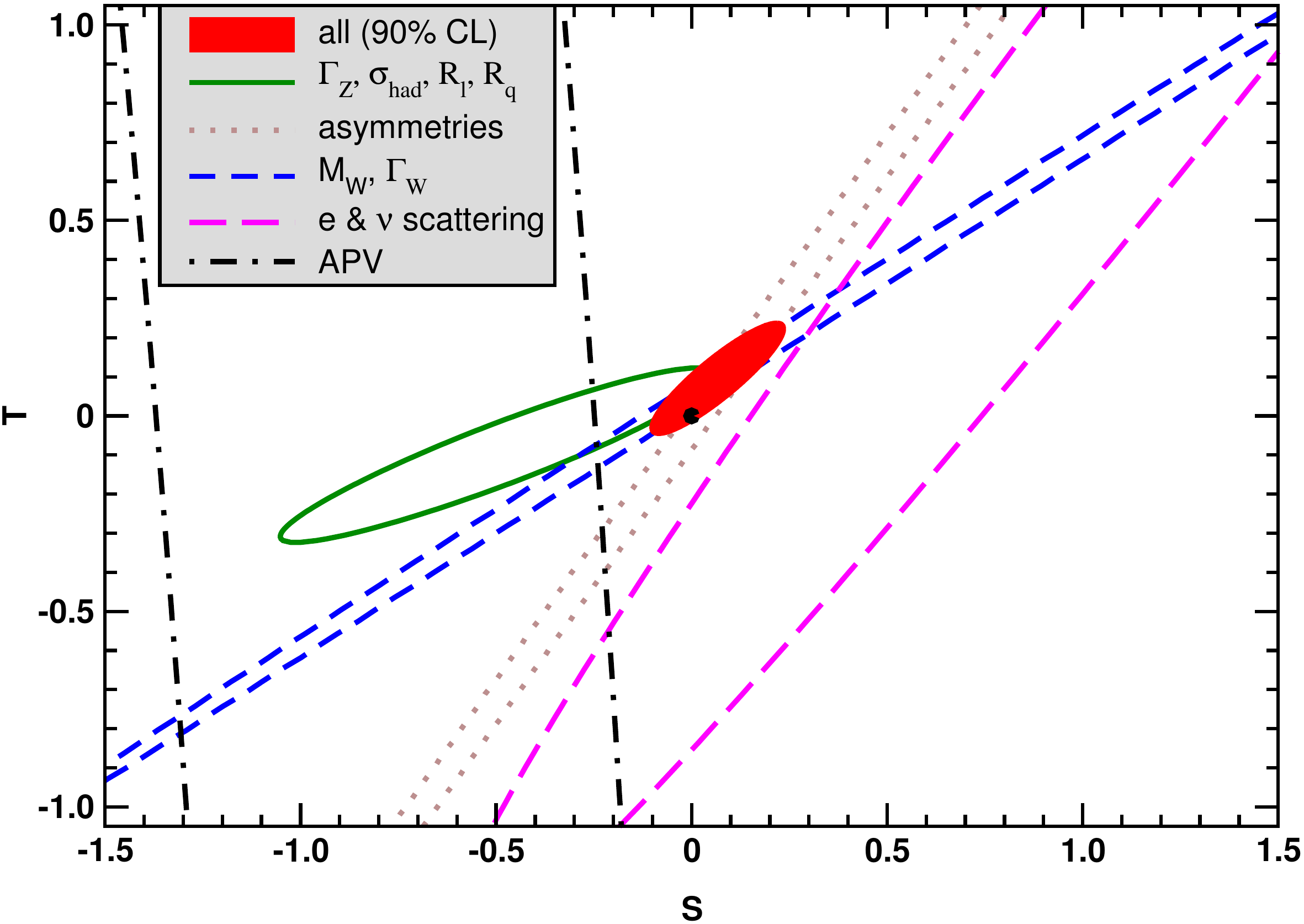}}
\caption[]{One standard deviation constraints on $S$ and $T$ from various data sets and the fit to all data~\cite{PDG2016}.}
\label{fig:ST}
\end{figure}

\section{Constraints on physics beyond the SM}
\subsection{Oblique physics beyond the SM}
\label{sec:oblique}
The oblique parameters, $S$, $T$ and $U$, describe corrections to the $W$ and $Z$ boson self-energies.
The SM contributions are subtracted out by definition, so that $S$, $T$ and $U$ are new physics parameters, where 
$S$ and $T$ (see in Fig.~\ref{fig:ST}) correspond to dimension six operators in the effective field theory, and $U$ is of dimension eight.
$T$ breaks the custodial $SO(4)$ symmetry of the Higgs potential. 
A multiplet of heavy {\em degenerate\/} chiral fermions contributes a fixed amount to $S$,
\begin{equation} 
\Delta S = \frac{N_C}{3\pi} \sum_i (t_{3L}^i - t_{3R}^i)^2,
\end{equation} 
where $t_{3L}$ and $ t_{3R}$ are the third components of weak isospin of the extra left and right-handed fermions, respectively.
Thus, for example, an additional {\em degenerate\/} fermion family yields
\begin{equation} 
\Delta S = \frac{2}{3\pi} \approx 0.21
\end{equation} 
The updated EW fit with $S$ and $T$ allowed simultaneously gives a range of values
\be
S = 0.06 \pm 0.08 \qquad\qquad 
T = 0.09 \pm 0.06 \qquad\qquad
\Delta\chi^2 = - 4.0
\ee
which are in marginal agreement with the SM but the decrease in $\chi^2$ relative to the SM is not insignificant.

\subsection{Implications of the T parameter}
The $T$ parameter has the same effect as the $\rho$ parameter --- the ratio of interaction strengths of the neutral and charged currents ---
as it is proportional to $\rho - 1$, but $T$ is often quoted for loop effects.
The $\rho$ parameter constrains vacuum expectation values of higher dimensional Higgs representations to $\lesssim 1$~GeV.
There is also sensitivity to {\em degenerate\/} scalar doublets up to 2~TeV,
a result based on an effective field theory approach~\cite{Henning:2014wua}.

Most importantly, non-degenerate doublets of additional fermions or scalars contribute an amount~\cite{Veltman:1977kh},
\be
\Delta\rho = \frac{G_F}{\sqrt{2}} \sum_i \frac{C_i}{8 \pi^2} \Delta m_i^2 \qquad\qquad
\Delta m_i^2 \geq (m_1 - m_2)^2,
\ee
where $C_i$ is the color factor.
$\Delta m_i^2$ is not exactly $m_1^2 - m_2^2$, where $m_i$ are the masses of the two members of the doublet,
but is a more complicated function bounded by $(m_1 - m_2)^2$ and thus gives rise to a positive-definite contribution. 
Despite the appearance of this form which seems to suggest that there is sensitivity to mass splittings 
even when the $m_i$ increase all the way to the Planck scale, {\em there is\/} decoupling of these heavy fermions or scalars,
because in models one will always face a see-saw type suppression of $\Delta m_i^2$ for very large $m_i$.
 
\begin{figure}[t]
\begin{minipage}{0.5\linewidth}
\centerline{\includegraphics[width=1.0\linewidth]{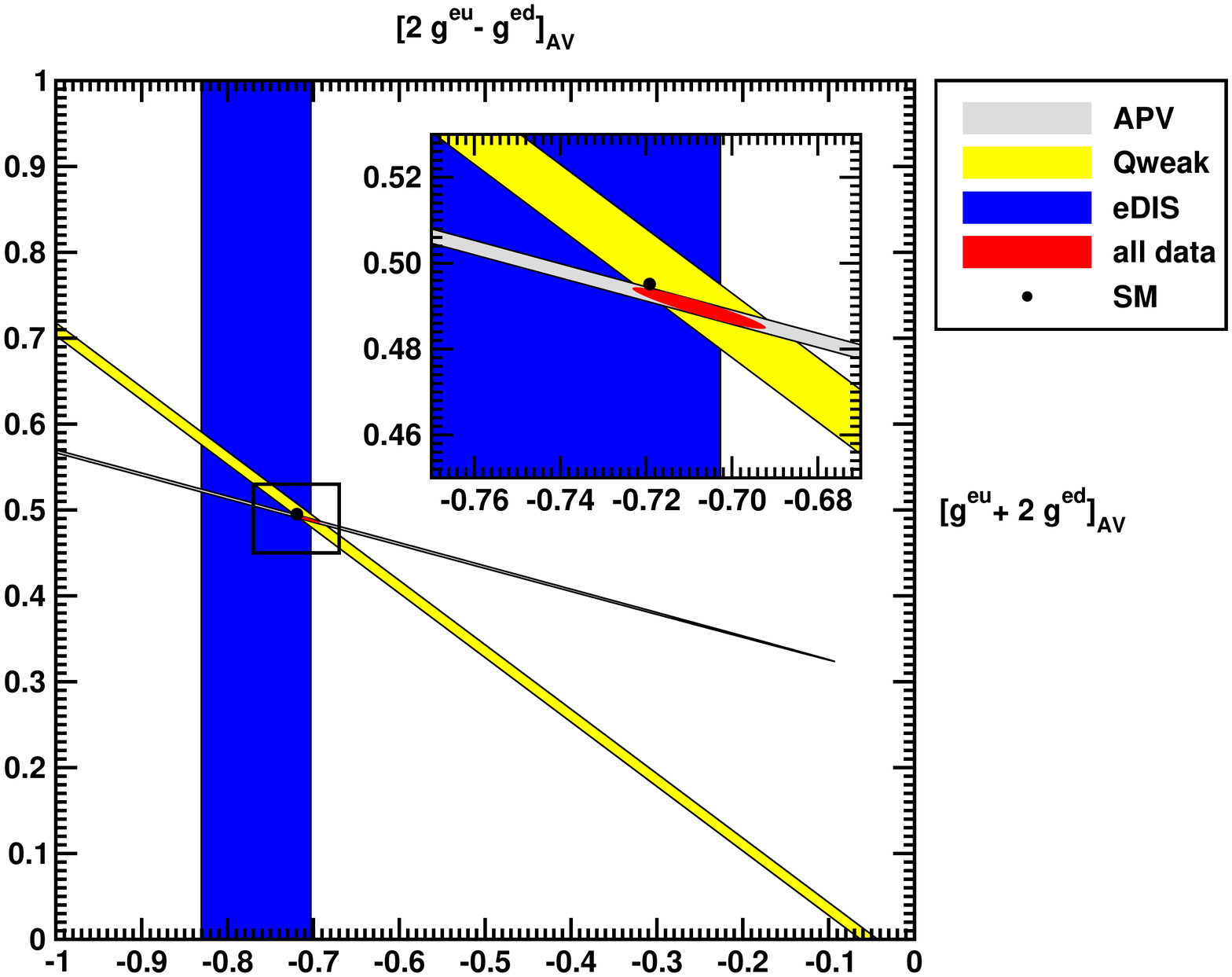}}
\end{minipage}
\hfill
\begin{minipage}{0.5\linewidth}
\centerline{\includegraphics[width=1.0\linewidth]{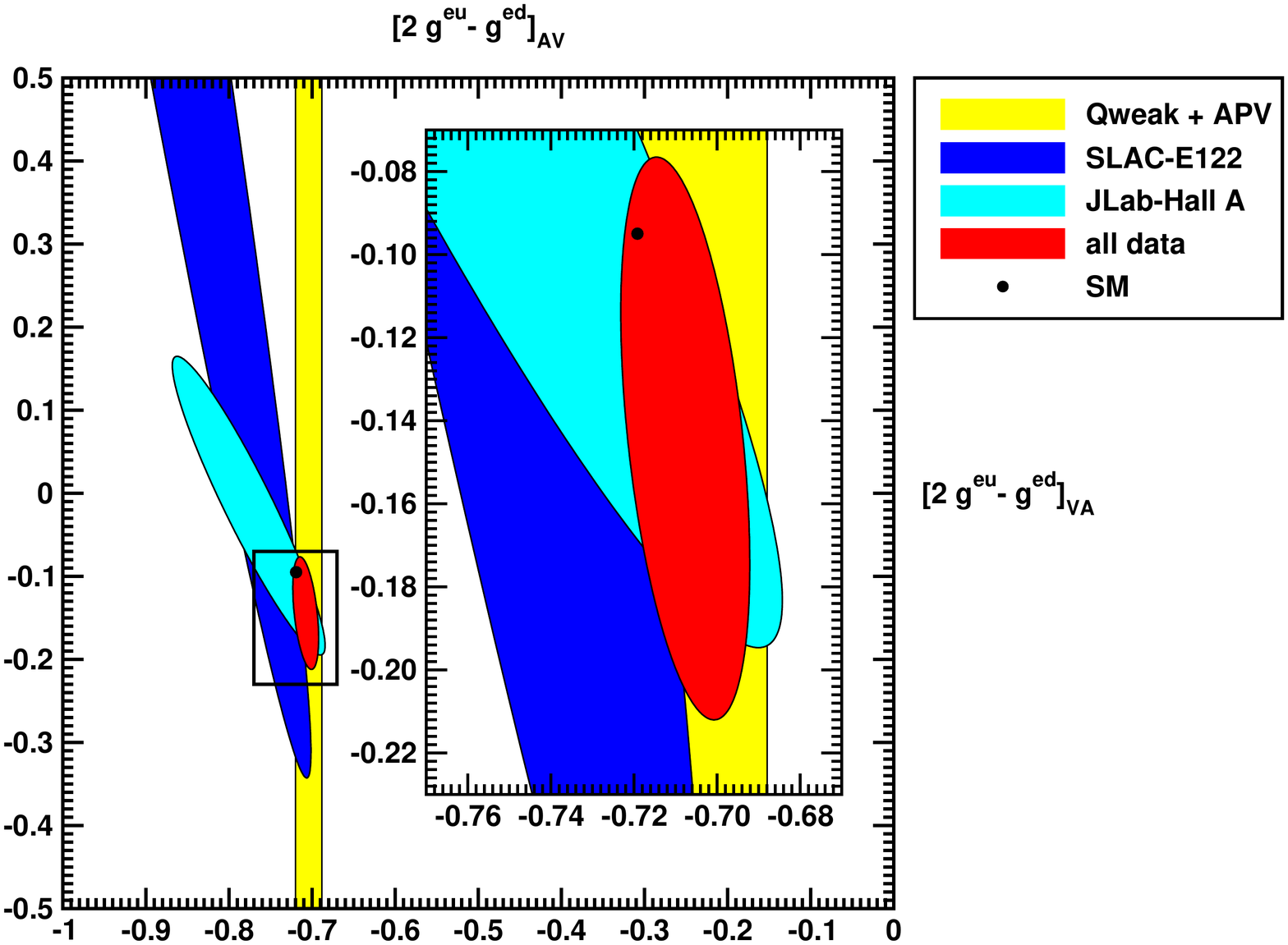}}
\end{minipage}
\caption[]{Experimentally determined axial-electron and vector-quark coupling combination $2 g_{AV}^{eu} - g_{AV}^{ed}$
{\em vs.\/} the orthogonal combination (left) and the vector-electron and axial-quark combination 
$2 g_{VA}^{eu} - g_{VA}^{ed}$ (right)~\cite{Erler:2014fqa}.}
\label{fig:couplings}
\end{figure}

I updated the one-parameter fit --- just allowing $\rho$ (or $T$) in addition to the SM parameters ---
with the result that $\rho$ is now 1.9~$\sigma$ above the SM prediction of unity,
\be
\rho = 1.00036 \pm 0.00019,
\ee
and thus one can constrain the sum of contributions of any additional EW doublet,
\be
\sum_i \frac{C_i}{3} \Delta m_i^2 \leq (46 \mbox{ GeV})^2 \qquad\qquad (95\% \mbox{ CL}).
\ee
Looking ahead, the LHC after the accumulation of 150~fb$^{-1}$ of data (with the same assumptions as in Sec.~\ref{sec:mh})
could reduce the error in $\rho$ to imply a stronger constraint on the mass splittings,
\be
\rho = 1 \pm 0.00014 \qquad\qquad \Longrightarrow \qquad\qquad
\sum_i \frac{C_i}{3} \Delta m_i^2 \leq (27 \mbox{ GeV})^2.
\ee
Or assuming that there is no change in the central value from today, one would actually obtain a precise measurement of $\Delta m_i^2$,
\be
\rho = 1.00036 \pm 0.00014 \qquad\qquad \Longrightarrow \qquad\qquad 
\sum_i \frac{C_i}{3} \Delta m_i^2 = (34^{+6}_{-7} \mbox{ GeV})^2.
\ee
Finally, turning to the HL--LHC one would find even stronger constraints,
\be
\rho = 1 \pm 0.00012 \qquad\qquad \Longrightarrow \qquad\qquad
\sum_i \frac{C_i}{3} \Delta m_i^2 \leq (25 \mbox{ GeV})^2.
\ee

\begin{figure}[t]
\centerline{\includegraphics[width=0.7\linewidth]{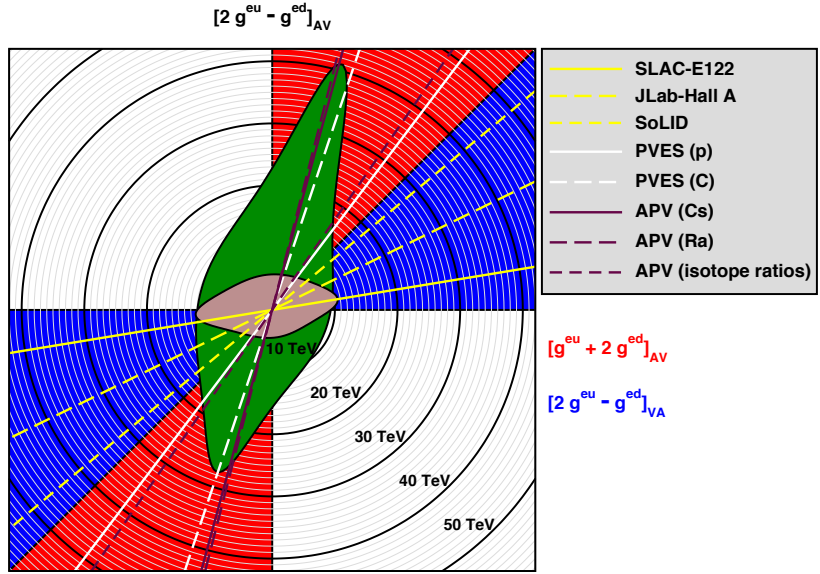}}
\caption[]{Compositeness scales~\cite{Erler:2014fqa} corresponding to the couplings in Fig.~\ref{fig:couplings}. 
They can be displayed as two overlaid planes since the horizontal axes coincide.
The lines define the coupling combinations tested by various types of experiment.
The blue segment is accessible to PVDIS~\cite{Souder:2016xcn,Wang:2014bba} (yellow lines) 
and defines the plane containing the brown 95\% CL exclusion contour.
Perpendicular to this is the plane containing the green contour and the red segment 
accessible to elastic polarized electron scattering (white lines) and APV (maroon lines).}
\label{fig:scales}
\end{figure}

\subsection{Compositeness scales from low energies}
Returning to the contact interactions in Sec.~\ref{sec:LE}
that may be derived by comparing on-pole and off-pole measurements of $\sin^2\theta_W$,
Fig.~\ref{fig:couplings} shows constraints on effective couplings corresponding to various parity-violating
effective four-fermion operators (as before, the couplings are defined to vanish in the absence of new physics).
These can be translated into compositeness scales that can be tested~\footnote{The numerical values of such scales
are convention dependent.  We use those laid out in Ref.~\cite{Erler:2013xha}.}.
As shown in Fig.~\ref{fig:scales} the new physics reach already surpassed 40~TeV and will increase beyond 50~TeV 
when the future experimental results from polarized electron scattering briefly mentioned in Sec.~\ref{sec:LE} 
are combined with measurements of atomic parity violation (APV).

\section{Conclusions}
The SM remains in remarkable health.
It is over-constrained, as $M_W$, $\sin^2\theta_W$, $g_\mu-2$, and many other quantities 
have been simultaneously computed and measured. 
If there is strongly coupled new physics its energy scale can be tested up to ${\cal O}(50~{\rm TeV})$ 
through parity-violating four-fermion operators. 

There are some inconclusive, yet interesting deviations.
$M_H$ extracted from the EW fit is $1.7~\sigma$ below the direct value,
and it is therefore mandatory to increase the precision in $m_t$ further and to obtain mutual consistency among different experiments. 
In a one-parameter fit ($S= U = 0$) the $T$ parameter appears $1.9~\sigma$ high,
and future measurements of $M_W$ at the LHC may increase this to a $3~\sigma$ effect.
Given that $M_W$ is particularly sensitive to physics beyond the SM and theoretically clean,
one may argue that a deviation in $M_W$ may be even more tantalizing than the current $4~\sigma$ SM discrepancy in $g_\mu-2$.
Thus, greater precision in $M_W$ is a must, with or or without an LHC discovery.

%All files (.tex, figures and .pdf) should be sent by the {\bf 15th of May 2017} by e-mail to {\bf moriond@in2p3.fr}.\\

\section*{Acknowledgments}
I would like to thank the organizers for the kind invitation to a very enjoyable meeting in a beautiful location.
This work is supported by CONACyT (M\'exico) project 252167--F.


\begin{thebibliography}{99}
\bibitem{Zupan} Jure Zupan, these proceedings.

\bibitem{Bouchendira:2010es} 
  R.~Bouchendira {\it et al.},
  %``New determination of the fine structure constant and test of the quantum electrodynamics,''
  {\em Phys.\ Rev.\ Lett.}\  {\bf 106}, 080801 (2011).
  %doi:10.1103/PhysRevLett.106.080801
  %arXiv:1012.3627 [physics.atom-ph].
  %%CITATION = doi:10.1103/PhysRevLett.106.080801;%%

\bibitem{Webber:2010zf} 
  MuLan Collaboration: D.~M.~Webber {\it et al.},
  %``Measurement of the Positive Muon Lifetime and Determination of the Fermi Constant to Part-per-Million Precision,''
  {\em Phys.\ Rev.\ Lett.}\  {\bf 106}, 041803 (2011).
  %[Phys.\ Rev.\ Lett.\  {\bf 106}, 079901 (2011)]
  %doi:10.1103/PhysRevLett.106.041803, 10.1103/PhysRevLett.106.079901
  %arXiv:1010.0991 [hep-ex].
  %%CITATION = doi:10.1103/PhysRevLett.106.041803, 10.1103/PhysRevLett.106.079901;%%
  
\bibitem{PDG2016} J.~Erler and A.~Freitas, {\em Electroweak Model and Constraints on New Physics}, in Ref.~\cite{Olive:2016xmw}.

\bibitem{Olive:2016xmw} 
  Particle Data Group: C.~Patrignani {\it et al.},
  %``Review of Particle Physics,''
  {\em Chin.\ Phys.}\ C {\bf 40}, 100001 (2016).
  %doi:10.1088/1674-1137/40/10/100001
  %%CITATION = doi:10.1088/1674-1137/40/10/100001;%%

\bibitem{ALEPH:2005ab} 
  ALEPH, DELPHI, L3, OPAL and SLD Collaborations, LEP EW Working Group and \\
  SLD EW and Heavy Flavour Groups: S.~Schael {\it et al.},
  %``Precision electroweak measurements on the $Z$ resonance,''
  {\em Phys.\ Rept.}\  {\bf 427}, 257 (2006).
  %doi:10.1016/j.physrep.2005.12.006
  %hep-ex/0509008.
  %%CITATION = doi:10.1016/j.physrep.2005.12.006;%%

\bibitem{Oda} Susumu Oda, these proceedings.

\bibitem{Owen} Mark Owen, these proceedings.

\bibitem{Bartos} Pavol Barto$\check{\rm s}$, these proceedings.

\bibitem{Erler:2015nsa} 
  J.~Erler,
  %``On the Combination Procedure of Correlated Errors,''
  {\em Eur.\ Phys.\ J.}\ C {\bf 75}, 453 (2015).
  %doi:10.1140/epjc/s10052-015-3688-y
  %arXiv:1507.08210 [physics.data-an].
  %%CITATION = doi:10.1140/epjc/s10052-015-3688-y;%%

\bibitem{Beneke:2016cbu} 
  M.~Beneke, P.~Marquard, P.~Nason and M.~Steinhauser,
  %``On the ultimate uncertainty of the top quark pole mass,''
  arXiv:1605.03609 [hep-ph].
  %%CITATION = ARXIV:1605.03609;%%

\bibitem{Erler:1998sy} 
  J.~Erler,
  %``Calculation of the QED coupling alpha (M(Z)) in the modified minimal subtraction scheme,''
  {\em Phys.\ Rev.}\ D {\bf 59}, 054008 (1999).
  %doi:10.1103/PhysRevD.59.054008
  %hep-ph/9803453.
  %%CITATION = doi:10.1103/PhysRevD.59.054008;%%

\bibitem{Erler:2004in} 
  J.~Erler and M.~J.~Ramsey-Musolf,
  %``The Weak mixing angle at low energies,''
  {\em Phys.\ Rev.}\ {\bf D 72}, 073003 (2005).
  %doi:10.1103/PhysRevD.72.073003
  %hep-ph/0409169.
  %%CITATION = doi:10.1103/PhysRevD.72.073003;%%

\bibitem{Erler:2000nx} 
  J.~Erler and M.~Luo,
  %``Hadronic loop corrections to the muon anomalous magnetic moment,''
  {\em Phys.\ Rev.\ Lett.}\ {\bf 87}, 071804 (2001).
  %doi:10.1103/PhysRevLett.87.071804
  %hep-ph/0101010.
  %%CITATION = doi:10.1103/PhysRevLett.87.071804;%%

\bibitem{Jegerlehner:2011ti} 
  F.~Jegerlehner and R.~Szafron,
  %``$\rho^0 - \gamma$ mixing in the neutral channel pion form factor $F_{\pi}^{e}$ and its role in comparing $e^+ e^-$ with $\tau$ spectral 
  %  functions,''
  {\em Eur.\ Phys.\ J.}\ C {\bf 71}, 1632 (2011).
  %doi:10.1140/epjc/s10052-011-1632-3
  %arXiv:1101.2872 [hep-ph].
  %%CITATION = doi:10.1140/epjc/s10052-011-1632-3;%%

\bibitem{Haisch} Uli Haisch, these proceedings.

\bibitem{Erler:2016atg} 
  J.~Erler, P.~Masjuan and H.~Spiesberger,
  %``Charm Quark Mass with Calibrated Uncertainty,''
  {\em Eur.\ Phys.\ J.}\ C {\bf 77}, 99 (2017).
  %doi:10.1140/epjc/s10052-017-4667-2
  %arXiv:1610.08531 [hep-ph].
  %%CITATION = doi:10.1140/epjc/s10052-017-4667-2;%%

\bibitem{Erler:2009jh} 
  J.~Erler, P.~Langacker, S.~Munir and E.~Rojas,
  %``Improved Constraints on Z-prime Bosons from Electroweak Precision Data,''
  {\em JHEP} {\bf 0908}, 017 (2009).
  %doi:10.1088/1126-6708/2009/08/017
  %arXiv:0906.2435 [hep-ph].
  %%CITATION = doi:10.1088/1126-6708/2009/08/017;%%
  
\bibitem{Andari} Nansi Andari, these proceedings.

\bibitem{Han} Liang Han, these proceedings.

\bibitem{Benesch:2014bas} 
  MOLLER Collaboration: J.~Benesch {\it et al.},
  %``The MOLLER Experiment: An Ultra-Precise Measurement of the Weak Mixing Angle Using M{\o}ller  
  %  Scattering,''
  arXiv:1411.4088 [nucl-ex].
  %%CITATION = ARXIV:1411.4088;%%
  
\bibitem{Androic:2013rhu} 
  Qweak Collaboration: D.~Androic {\it et al.},
  %``First Determination of the Weak Charge of the Proton,''
  {\em Phys.\ Rev.\ Lett.}\ {\bf 111}, 141803 (2013).
  %doi:10.1103/PhysRevLett.111.141803
  %arXiv:1307.5275 [nucl-ex].
  %%CITATION = doi:10.1103/PhysRevLett.111.141803;%%

\bibitem{Bucoveanu:2016bgx} 
  R.~Bucoveanu, M.~Gorchtein and H.~Spiesberger, 
  %``Precision Measurement of $\sin^2 \theta_w$ at MESA,''
  PoS LL {\bf 2016}, 061 (2016).
  %arXiv:1606.09268 [hep-ph].
  %%CITATION = ARXIV:1606.09268;%%

\bibitem{Souder:2016xcn} 
  P.~A.~Souder,
  %``Parity Violation in Deep Inelastic Scattering with the SoLID Spectrometer at JLab,''
  {\em Int.\ J.\ Mod.\ Phys.\ Conf.\ Ser.}\ {\bf 40}, 1660077 (2016).
  %doi:10.1142/S2010194516600776
  %%CITATION = doi:10.1142/S2010194516600776;%%

\bibitem{Bodek:2015bya} 
  A.~Bodek,
  %``Standard Model Precision Electroweak Measurements at HL-LHC and Future Hadron Colliders,''
  arXiv:1510.02006 [hep-ex].
  %%CITATION = ARXIV:1510.02006;%%

\bibitem{Schael:2013ita} 
  ALEPH, DELPHI, L3 and OPAL Collaborations and LEP EW Working Group: \\ S.~Schael {\it et al.}, 
  %``Electroweak Measurements in Electron-Positron Collisions at W-Boson-Pair Energies at LEP,''
  {\em Phys.\ Rept.}\  {\bf 532}, 119 (2013).
  %doi:10.1016/j.physrep.2013.07.004
  %arXiv:1302.3415 [hep-ex].
  %%CITATION = doi:10.1016/j.physrep.2013.07.004;%%

\bibitem{Heinemeyer:2013dia} 
  S.~Heinemeyer, W.~Hollik, G.~Weiglein and L.~Zeune,
  %``Implications of LHC search results on the W boson mass prediction in the MSSM,''
  {\em JHEP} {\bf 1312}, 084 (2013).
  %doi:10.1007/JHEP12(2013)084
  %arXiv:1311.1663 [hep-ph].
  %%CITATION = doi:10.1007/JHEP12(2013)084;%%

\bibitem{Baur:2002gp} 
  Snowmass Working Group on Precision EW Measurements: U.~Baur {\it et al.},
  %``Present and future electroweak precision measurements and the indirect determination of the mass of the Higgs boson,''
  %{\em eConf} {\bf C010630}, P1WG1 (2001).
  hep-ph/0202001.
  %%CITATION = HEP-PH/0202001;%%
  
\bibitem{Bozzi:2015zja} 
  G.~Bozzi, L.~Citelli, M.~Vesterinen and A.~Vicini,
  %``Prospects for improving the LHC W boson mass measurement with forward muons,''
  {\em Eur.\ Phys.\ J}.\ C {\bf 75}, 601 (2015).
  %doi:10.1140/epjc/s10052-015-3810-1
  %arXiv:1508.06954 [hep-ex].
  %%CITATION = doi:10.1140/epjc/s10052-015-3810-1;%%
  
\bibitem{Henning:2014wua} 
  B.~Henning, X.~Lu and H.~Murayama,
  %``How to use the Standard Model effective field theory,''
  {\em JHEP} {\bf 1601}, 023 (2016).
  %doi:10.1007/JHEP01(2016)023
  %arXiv:1412.1837 [hep-ph].
  %%CITATION = doi:10.1007/JHEP01(2016)023;%%
  
\bibitem{Veltman:1977kh} 
  M.~J.~G.~Veltman,
  %``Limit on Mass Differences in the Weinberg Model,''
  {\em Nucl.\ Phys.}\ B {\bf 123}, 89 (1977).
  %doi:10.1016/0550-3213(77)90342-X
  %%CITATION = doi:10.1016/0550-3213(77)90342-X;%%

\bibitem{Erler:2013xha} 
  J.~Erler and S.~Su,
  %``The Weak Neutral Current,''
  {\em Prog.\ Part.\ Nucl.\ Phys.}\  {\bf 71}, 119 (2013).
  %doi:10.1016/j.ppnp.2013.03.004
  %arXiv:1303.5522 [hep-ph].
  %%CITATION = doi:10.1016/j.ppnp.2013.03.004;%%

\bibitem{Erler:2014fqa} 
  J.~Erler, C.~J.~Horowitz, S.~Mantry and P.~A.~Souder, \\
  %``Weak Polarized Electron Scattering,''
  {\em Ann.\ Rev.\ Nucl.\ Part.\ Sci.}\  {\bf 64}, 269 (2014).
  %doi:10.1146/annurev-nucl-102313-025520
  %arXiv:1401.6199 [hep-ph].
  %%CITATION = doi:10.1146/annurev-nucl-102313-025520;%%

\bibitem{Wang:2014bba} 
  PVDIS Collaboration: D.~Wang {\it et al.},
  %``Measurement of parity violation in electron\UTF{00D0}quark scattering,''
  Nature {\bf 506}, 67 (2014).
  %doi:10.1038/nature12964
  %%CITATION = doi:10.1038/nature12964;%%  
  
\end{thebibliography}
\end{document}